\documentclass[journal,table]{IEEEtran}

\ifCLASSINFOpdf
\else
   \usepackage[dvips]{graphicx}
\fi
\usepackage{url}
\usepackage{threeparttable}
\usepackage{pifont}
\usepackage{orcidlink}
\usepackage{amsmath, amssymb}
\usepackage{pifont}
\usepackage{makecell}
\usepackage{graphicx} 
\usepackage{booktabs} 
\usepackage{multirow} 
\usepackage{amsmath}  
\usepackage{tabularx}  
\usepackage{colortbl}
\usepackage[section]{placeins}
\usepackage{array}    
\usepackage{adjustbox} 

\begin{document}

\title{Large Language Models Can Achieve \\Explainable and Training-Free One-shot HRRP ATR}

\author{Lingfeng Chen$^{\orcidlink{0009-0003-2690-9407}}$, Panhe Hu$^{\orcidlink{0009-0003-2690-9407}}$, Zhiliang Pan, Qi Liu, and Zhen Liu$^{\orcidlink{0000-0002-1233-1494}}$, \IEEEmembership{Member, IEEE}
\vspace{-2em}
\thanks{Manuscript received June 24, 2025; revised XXX, 2025; accepted XXX, 2025. Date of publication XXX, 2025; date of current
version XXX, 2025. This work was supported in part by National Natural Science Foundation of China under Grant 62201588, in part by Natural Science Foundation of Hunan under Grant 2024JJ10007. The associate editor coordinating the review of this article and approving it for publication was XXX. \textit{(Corresponding author:
Panhe Hu
.)}}
\thanks{Lingfeng Chen, Panhe Hu, Zhiliang Pan, Qi Liu, and Zhen Liu
are with the College of Electronic Science and Technology,
National University of Defense Technology, Changsha 410073, Hunan, China (e-mail: \href{chenlingfeng@nudt.edu.cn}{chenlingfeng@nudt.edu.cn}; \href{hupanhe13@nudt.edu.cn}{hupanhe13@nudt.edu.cn}; \href{panzhiliang17@nudt.edu.cn}{panzhiliang17@nudt.edu.cn}; \href{ouc_liuqi@163.com}{ouc\_liuqi@163.com}; \href{zhen_liu@nudt.edu.cn}{zhen\_liu@nudt.edu.cn}}

\thanks{Digital Object Identifier XX.XXXX/XXXXX.2025.XXXXXX}}

\markboth{SUMITTED TO IEEE SIGNAL PROCESSING LETTERS}
{Chen \MakeLowercase{\textit{et al.}}: Explainable and Training-Free Radar HRRP Target Recognition via In-Context Learning with Large Language Models}
\maketitle

\begin{abstract}
This letter introduces a pioneering, training-free and explainable framework for High-Resolution Range Profile (HRRP) automatic target recognition (ATR) utilizing large-scale pre-trained Large Language Models (LLMs). Diverging from conventional methods requiring extensive task-specific training or fine-tuning, our approach converts one-dimensional HRRP signals into textual scattering center representations. Prompts are designed to align LLMs' semantic space for ATR via few-shot in-context learning, effectively leveraging its vast pre-existing knowledge without any parameter update. We make our codes publicly available at \href{https://uithub.com/MountainChenCad/HRRPLLM}{https://uithub.com/MountainChenCad/HRRPLLM} to foster research into LLMs for HRRP ATR.
\end{abstract}


\IEEEpeerreviewmaketitle

\section{Introduction}

\begin{figure}[t]
\centerline{\includegraphics[width=\linewidth]{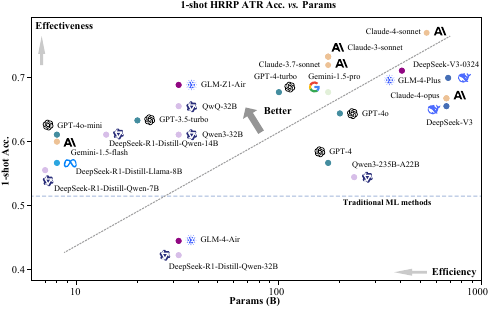}}
\caption{Averaged 3-way 1-shot accuracy on simulated 12 types HRRP dataset. Key findings: (1) LLMs simply yield good performance over current baselines; (2) ATR performance emergent as the scale of LLM increases; (3) distilled smaller LLMs can also achieve competitive performance. For models without public scale information, we inferred the approximate parameter count.}
\end{figure}

\begin{figure}[t]
\centerline{\includegraphics[width=0.85\linewidth]{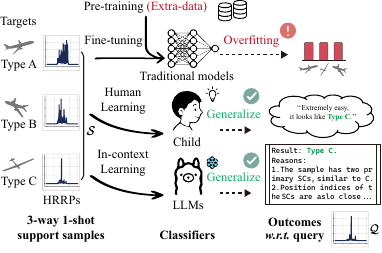}}
\caption{Traditional models fails to generalize on few-shot samples, while human (even a child) can sometimes easily guess the type of HRRP targets. Based on our HRRPLLM framework, LLMs can also generalize easily on novel types without extra-training and explain the reason of its recognition.}
\end{figure}                                                 

\IEEEPARstart{H}{igh}-Resolution Range Profiles (HRRPs), \textit{i.e.}, 1-D radar line-of-sight (LOS) projections of target scattering centers (SCs)~\cite{liu2024prior}, are effective structural signatures for automatic target recognition (ATR) task~\cite{chen2024hrrpgraphnet, yang2024radar, wang2023type, li2024priorinfo, song2022multiview, liu2021fewshot, guo2025hrrp, li2023fewshotigarss, liu2025attribute, liu2023contrastive, liu2024scnet, liu2021gafmn, liu2025radarhybrid}. Deep learning (DL) based HRRP ATR methods surpass traditional methods through efficient feature representation~\cite{xia2023radaropen, liu2021multipolarization, li2025mtbc}. However, traditional DL models struggle with generalization in few-shot scenarios~\cite{liu2021fewshot}.

Few-shot learning (FSL) aims to address this by enabling models to learn from minimal examples like humans~\cite{kumar_meta-learning_2024}. Current HRRP FSL methods (\textit{e.g.}, meta-learning~\cite{liu2021fewshot}, metric-learning~\cite{li2023fewshotigarss}, data augmentation \cite{song2022multiview}) face limitations. \textbf{First,} their generalization to novel classes is restricted \cite{wolpert1997nofree, guo2025hrrp}. \textbf{Second,} they can be computationally expensive~\cite{li2025mtbc}. \textbf{Third,} recognition results explainability remains a challenge~\cite{sagayaraj2021comparative}.

Large Language Models (LLMs) have shown remarkable FSL capabilities through in-context learning (ICL) across many domains~\cite{wang2024tabletime, dong2024survey, liu2022whatmakes}. While vision~\cite{radford2021learning} and remote sensing~\cite{liu2024remoteclip, li2025saratrx, liu2024infmae} have seen foundation model development, the HRRP field has not, primarily due to lack of publicly available datasets~\cite{song2022multiview}, and the modality gap~\cite{li2024sardet100k}. Despite HRRPs being 1-D projections and thus sparser in information content compared to 2-D images, its sample structure naturally fits LLMs' text modality~\cite{gruver2023large, tang2025time, wang2024tabletime, jiang2024empowering}. This motivates us: \textit{Can general-purpose LLMs, through ICL and appropriate HRRP feature textualization, perform effective few-shot HRRP target recognition without any task-specific training?}


This letter introduces \textbf{HRRPLLM}, a novel framework enabling, to our knowledge, the first explainable and training-free HRRP ATR via LLMs. \textbf{Our core innovations are twofold:} \textbf{\ding{182}} We managed to bridge the modality gap between electromagnetic HRRPs and the LLMs' semantic space by extracting SCs and textual serialization. \textbf{\ding{183}} We design a sophisticated prompting strategy that leverages ICL for both ATR and human-understandable explanations. Remarkably, HRRPLLM achieves competitive 1-shot results against HRRP ATR baselines on both simulated and measured datasets.

\section{Methodology}


\subsection{Problem Setup}
Few-shot HRRP ATR is $N$-way $K$-shot classification on $N_\mathcal{Q}$ queries. A FSL task $\mathcal{T}$ is constructed by $N_\mathcal{C}$ distinct target classes. For each of these $N_\mathcal{C}$ classes, $N_\mathcal{S}$ labeled support samples are selected to form the support set $\mathcal{D}_{\mathcal{S}}^{\mathcal{T}} = \{ (\mathbf{X}_i^{(\mathcal{S})}, y_i^{(\mathcal{S})}) \}_{i=1}^{N_\mathcal{C} \times N_\mathcal{S}}$, where $\mathbf{X}_i^{(\mathcal{S})}$ is the $i$-th support HRRP and $y_i^{(\mathcal{S})}$ is its corresponding class label. Additionally, for each of the $N_\mathcal{C}$ classes, $N_\mathcal{Q}$ distinct samples are selected to form the query set $\mathcal{D}_{\mathcal{Q}}^{\mathcal{T}} = \{ (\mathbf{X}_j^{(\mathcal{Q})}, y_j^{(\mathcal{Q})}) \}_{j=1}^{N_\mathcal{C} \times N_\mathcal{Q}}$. Model $\Theta$ predicts the labels $y_j^{(\mathcal{Q})}$ for the query samples $\mathbf{X}_j^{(\mathcal{Q})}$, based on support set $\mathcal{D}_{\mathcal{S}}^{\mathcal{T}}$. That is learning a mapping $f_\Theta({\mathcal{T}}): \mathbf{X}^{(\mathcal{Q})} \rightarrow \hat{y}^{(\mathcal{Q})}$ conditioned on $\mathcal{D}_{\mathcal{S}}^{\mathcal{T}}$, where $\hat{y}^{(\mathcal{Q})}$ is the predicted label.

\begin{figure}[t]
    \centering
    \includegraphics[width=\linewidth]{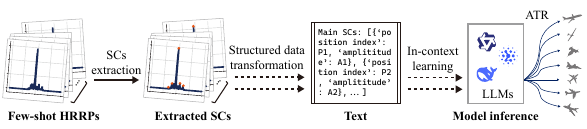}
    \caption{The general workflow of the proposed HRRPLLM framework. In a word, our framework transforms few-shot HRRP ATR tasks to ICL tasks for LLMs reason by generating SCs-based prompts. 
    }
    \label{fig:method_example}
\end{figure}

\subsection{HRRP Signal Formation}
A HRRP is a discrete complex-valued signal $\mathbf{X} \in \mathbb{C}^{N_r}$, where $N_r \in \mathbb{N}^+$ is the number of range resolution cells, or bins. Each element $x(k)$ of $\mathbf{X}$, for $k \in \{0, \dots, N_r-1\}$, represents the coherently processed radar echo intensity corresponding to the $k$-th range bin, $r_k = k \Delta R$, where $\Delta R$ is the radar's range resolution. The HRRP $\mathbf{X}$ is synthesized from wideband radar returns. Consider a transmitted Linear Frequency Modulated (LFM) pulse $s_{tx}(t; f_c, B, T_p)$, characterized by center frequency $f_c$, bandwidth $B$, and pulse duration $T_p$. The received signal $s_{rx}(t)$ is processed via involving dechirping and matched filtering to yield a discrete frequency domain representation $\mathbf{R} \in \mathbb{C}^{M}$. Each component $R(m)$ ($m \in \{0, \dots, M-1\}$) of $\mathbf{R}$ corresponds to the complex echo amplitude at a discrete frequency $f_m = f_{start} + m \cdot (B/M)$. The transformation from the frequency domain $\mathbf{R}$ to the range domain $\mathbf{X} = \mathcal{F}^{-1}\{\mathbf{R}\}$ is achieved via an Inverse Discrete Fourier Transform (IDFT), calculated by:
\begin{align}
     x(k) = \frac{1}{\sqrt{M}} &\sum_{m=0}^{M-1} R(m) e^{j 2\pi \frac{mk}{M}}, \quad \forall k \in \mathcal{K}_r,
    \label{eq:hrrp_sum_detail}
\end{align}
where $\mathcal{K}_r = \{0, \dots, N_r-1\}$ and $N_r=M$. The scaling $1/\sqrt{M}$ ensures a unitary transform. The resulting $\mathbf{X}$ captures the target's SCs distribution along the radar LOS \cite{chen2024hrrpgraphnet}. The magnitude $|x(k)|$ is particularly indicative of scattering phenomena \cite{liu2024prior}. The range resolution is $\Delta R = c/(2B)$.

\subsection{Dominant Scattering Center Extraction}
\label{sec:sc}
To obtain a sparse and physically salient representation from an HRRP $\mathbf{X} \in \mathbb{C}^{N_r}$, dominant SCs are extracted. Initially, the HRRP amplitude profile $\mathbf{A} \in \mathbb{R}^{N_r}_{\ge 0}$ is computed, where $A(k) = |x(k)|, \forall k \in \mathcal{K}_r$. Normalization of $\mathbf{A}$ by its $L_\infty$-norm yields $\mathbf{A}_{norm} \in [0,1]^{N_r}$. $A_{norm}(k) = A(k) / \|\mathbf{A}\|_{\infty}$, where $ \|\mathbf{A}\|_{\infty} = \max_{k' \in \mathcal{K}_r} A(k')$. Peaks in $\mathbf{A}_{norm}$ are identified based on criteria involving amplitude, prominence $P(\mathbf{A}_{norm}, k)$~\cite{liu2024scnet}, and minimum separation $d_{th}$. Let $\mathcal{L}(\mathbf{A}_{norm})$ be the set of indices of local maxima. The set of qualifying peak locations $\mathcal{P}_{idx}$ is:
\begin{align}
    \mathcal{P}_{idx} = \{ k \mid k \in \mathcal{L}(\mathbf{A}_{norm}) \land A_{norm}(k) > \tau_A \land \nonumber \\ 
    P(\mathbf{A}_{norm}, k) > \tau_p \land \min_{j \in \mathcal{P}_{idx} \setminus \{k\}} |r_k - r_j| > d_{th} \},
    \label{eq:peak_formal_criteria}
\end{align}
where $r_k=k\Delta R$, and $\tau_A, \tau_p$ are thresholds. The associated amplitudes are $a_k = A_{norm}(k)$ for $k \in \mathcal{P}_{idx}$.
Let $\mathcal{C}_{cand} = \{ (r_k, a_k) \mid k \in \mathcal{P}_{idx} \}$. An ordering relation $\succ_a$ is defined such that $(r_i, a_i) \succ_a (r_j, a_j)$ if $a_i > a_j$. The set of $M_{sc}$ dominant SCs for a single HRRP $\mathbf{X}$ is $\mathcal{S}_{\mathbf{X}}$:
\begin{align}
    \mathcal{S}_{\mathbf{X}} &= \{ (r_{(j)}, a_{(j)}) \}_{j=1}^{M_{sc}}, \nonumber\\\textbf{s.t. } (r_{(j)}, a_{(j)}) \text{ is}&~\text{the } j^{th} \text{ element in } \text{sort}(\mathcal{C}_{cand}, \succ_a).
    \label{eq:sc_set_ordered}
\end{align}


\subsection{Textual Serialization of Scattering Centers}
\label{sec:sctext}
For LLM ingestion, the numerical $\mathcal{S}_{\mathbf{X}} = \{ (r_j, a_j) \}_{j=1}^{M_{sc}}$ is simply serialized into a textual string $\text{Text}_{\mathcal{S}}$ via structured data encoding. \textit{E.g.}, in our prompt construction, we represent $\mathcal{S}$ as a list of dictionary-like structures, each $\texttt{\{'position~index:'}~r_j\texttt{,~} \texttt{'ampilitude:'}~a_j\texttt{\}}$ corresponding to an SC. 
This $\text{Text}_{\mathcal{S}}$ serves as the LLM input feature, preserving SC positional $r_j$ and amplitude $a_j$ data.

\subsection{Few-shot Recognition via In-Context Learning}
\label{sec:icl}
\begin{figure}[t]
    \centering
    \includegraphics[width=0.90\linewidth]{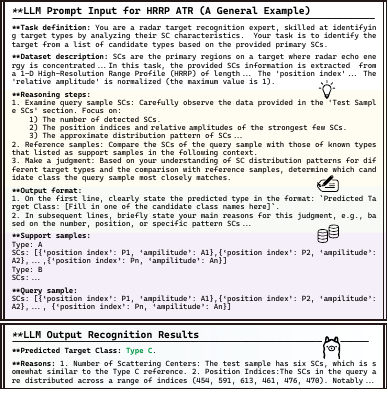}
    \caption{An general example structure of the prompt $\mathcal{P}_j^{\mathcal{T}}$ fed to the LLM (\textit{i.e.}, the contextual domain information, support \& query representation, and instructional cues) and a typical LLM output of our method.}
    \label{fig:prompt_example}
\end{figure}

Our framework employs LLMs, denoted $\mathcal{M}_{LLM}$, harnessing their ICL capability \cite{dong2024survey}. ICL allows $\mathcal{M}_{LLM}$ to perform a given few-shot task $\mathcal{T}$ by conditioning on a structured prompt $\mathcal{P}$, without any modification to the LLM's parameters $\theta_{LLM}$. Prompts $\mathcal{P}_j^{\mathcal{T}}$ is constructed \textit{w.r.t.} every query HRRP $\mathbf{X}_j^{(\mathcal{Q})}$ in a task $\mathcal{T}$. An example is illustrated in Fig.~\ref{fig:prompt_example}, this prompt meticulously assembles several key informational pieces to guide the LLMs: \textbf{(1) contextual domain information:} this includes a definition of the ATR task, descriptions of HRRPs and SCs, and a list of the $N_\mathcal{C}$ candidate target classes for the current task; \textbf{(2) reasoning steps:} several steps for efficient are  offered as primary instruction for LLMs' reasoning; \textbf{(3) output format:} this makes the results formatted and force the LLMs to speak out the detailed reasons for explainability; \textbf{(4) few-shot support samples and query:} support set $\mathcal{D}_{\mathcal{S}}^{\mathcal{T}}$ and consist of $N_{ex}$ pairs, where each pair comprises the textual SC representation of a support sample and its corresponding true class label. This set of support samples, $\mathcal{E}_{\mathcal{S}}^{\mathcal{T}}$, is denoted as $\mathcal{E}_{\mathcal{S}}^{\mathcal{T}} = \{ (\text{Text}_{\mathcal{S}_{ex,l}}, y_{ex,l}) \}_{l=1}^{N_{ex}}$ where $\text{Text}_{\mathcal{S}_{ex,l}}$ is the encoded SC string. Following this, textual SC representation of the query is provided, along with a clear instruction cue  $\mathcal{I}_{instr}$ directing the LLM to predict its class from the $N_\mathcal{C}$ candidates.\footnote[1]{We have also prepared a specific \href{https://uithub.com/MountainChenCad/HRRPLLM}{\textbf{HRRPLLM-DEMO Toy Example}} in our public repository for detailed illustration.}


\section{Experiments}

\begin{table}[ht] 
\centering
\caption{\textbf{Results of Comparative Few-shot HRRP ATR Experiments.\\} The LLMs in our HRRPLLM framework and the traditional\\ ML methods are evaluated on both simulated and\\ measured dataset. TF means Traning-Free.}
\label{tab:comparative_results}
\begin{adjustbox}{width=\linewidth,center} 
\begin{threeparttable}
\begin{tabular}{c c c c c c c c}

\cmidrule{1-8}
\textbf{Dataset}&\textbf{Type} & \textbf{Model}  & \textbf{Date} & \textbf{Params} & \textbf{TF} & \textbf{1-shot Acc.} \textbf{(}$\uparrow$\textbf{)} & \textbf{F1.} \textbf{(}$\uparrow$\textbf{)} \\
\cmidrule{1-8}
\multirow{28}{*}{\makecell{\textbf{Simulated}\\\textbf{HRRP}\\\textbf{Dataset}}} &\multirow{3}{*}{\makecell{\textbf{Traditional}\\\textbf{ML}}} 
& SVM-HRRP~\cite{svm}  & Sep. 2005& - & \ding{55} & 62.22  & 54.44 \\ 
&& SVM-SC~\cite{svm_sc}  & Dec. 2007 & - & \ding{55} & 51.11 & 43.89\\
&& RF-SC~\cite{rf}  & Aug. 2018 & - & \ding{55} & 54.44 & 47.22\\
\cmidrule{2-8}
\cmidrule{2-8}
&\multirow{15}{*}{\makecell{\textbf{API-based}\\\textbf{LLMs}\\\textbf{(Proposed)}}} 
& \cellcolor{green!20}{GPT-4o-mini}  & \cellcolor{green!20}Jul. 2024 & \cellcolor{green!20}8 B & \cellcolor{green!20}\ding{51} & \cellcolor{green!20}61.11 & \cellcolor{green!20}58.75
\\
&& \cellcolor{green!20}{GPT-3.5-turbo}  & \cellcolor{green!20}Mar. 2023 & \cellcolor{green!20}20 B & \cellcolor{green!20}\ding{51} & \cellcolor{green!20}63.33 & \cellcolor{green!20}61.15\\
&& \cellcolor{green!20}{GPT-4}  & \cellcolor{green!20}Apr. 2023 & \cellcolor{green!20}- & \cellcolor{green!20}\ding{51} & \cellcolor{green!20}56.67 & \cellcolor{green!20}51.50 \\
&& \cellcolor{green!20}{GPT-4-turbo}  & \cellcolor{green!20}Jan. 2024 & \cellcolor{green!20}- & \cellcolor{green!20}\ding{51} & \cellcolor{green!20}67.78 & \cellcolor{green!20}61.79\\
&& \cellcolor{green!60}{\textbf{GPT-o4-mini}}  & \cellcolor{green!60}\textbf{Apr. 2025} & \cellcolor{green!60}\textbf{-} & \cellcolor{green!60}\ding{51} & \cellcolor{green!60}\textbf{75.56} & \cellcolor{green!60}\textbf{75.40} \\
&& \cellcolor{green!20}{GPT-4.1}  & \cellcolor{green!20}Apr. 2025 & \cellcolor{green!20}- & \cellcolor{green!20}\ding{51} & \cellcolor{green!20}66.67 & \cellcolor{green!20}64.26\\
&& \cellcolor{green!60}{\textbf{Claude-3-sonnet}}  & \cellcolor{green!60}\textbf{Feb. 2024} & \cellcolor{green!60}\textbf{-} & \cellcolor{green!60}\ding{51} & \cellcolor{green!60}\underline{73.33} & \cellcolor{green!60}72.24\\
&& \cellcolor{green!20}{Claude-3-opus} & \cellcolor{green!20}Feb. 2024 & \cellcolor{green!20}- & \cellcolor{green!20}\ding{51} & \cellcolor{green!20}68.89 & \cellcolor{green!20}65.58\\
&& \cellcolor{green!20}{Claude-3-5-sonnet} & \cellcolor{green!20}Oct. 2024 & \cellcolor{green!20}- & \cellcolor{green!20}\ding{51} & \cellcolor{green!20}66.67 & \cellcolor{green!20}63.05\\
&& \cellcolor{green!20}{Claude-3-7-sonnet} & \cellcolor{green!20}Feb. 2025 & \cellcolor{green!20}- & \cellcolor{green!20}\ding{51} & \cellcolor{green!20}68.89 & \cellcolor{green!20}65.52\\
&& \cellcolor{green!60}{\textbf{Claude-4-sonnet}} & \cellcolor{green!60}\textbf{May 2025} & \cellcolor{green!60}\textbf{-} & \cellcolor{green!60}\ding{51} & \cellcolor{green!60}\textbf{75.56} & \cellcolor{green!60}\underline{71.82}\\
&& \cellcolor{green!20}{Claude-4-opus} & \cellcolor{green!20}May 2025 & \cellcolor{green!20}- & \cellcolor{green!20}\ding{51} & \cellcolor{green!20}66.67 & \cellcolor{green!20}62.16\\
&& \cellcolor{green!20}{Gemini-1.5-Pro} & \cellcolor{green!20}Sep. 2024 & \cellcolor{green!20}175 B & \cellcolor{green!20}\ding{51} & \cellcolor{green!20}71.11 & \cellcolor{green!20}70.04\\
&& {GLM-4-Air} & Jun. 2024 & - & \ding{51} & 42.22 & 38.79\\
&& \cellcolor{green!20}{GLM-4-Plus} & \cellcolor{green!20}Sep. 2024 & \cellcolor{green!20}- & \cellcolor{green!20}\ding{51} & \cellcolor{green!20}71.11 & \cellcolor{green!20}70.04\\
\cmidrule{2-8}
&\multirow{10}{*}{\makecell{\textbf{Open-source}\\\textbf{LLMs}\\\textbf{(Proposed)}}}  & \cellcolor{green!20}{DeepSeek-R1-Distill-Qwen-7B}  & \cellcolor{green!20}Jan. 2025 & \cellcolor{green!20}7 B & \cellcolor{green!20}\ding{51} & \cellcolor{green!20}55.56 & \cellcolor{green!20}54.30\\
&& \cellcolor{green!20}{DeepSeek-R1-Distill-Llama-8B}& \cellcolor{green!20}Jan. 2025 & \cellcolor{green!20}8 B & \cellcolor{green!20}\ding{51} & \cellcolor{green!20}56.67 & \cellcolor{green!20}55.26 \\
&& \cellcolor{green!20}{DeepSeek-R1-Distill-Qwen-14B}  & \cellcolor{green!20}Jan. 2025 & \cellcolor{green!20}14 B & \cellcolor{green!20}\ding{51} & \cellcolor{green!20}61.11 & \cellcolor{green!20}59.62\\
&& {DeepSeek-R1-Distill-Qwen-32B} & Jan. 2025 & 32 B & \ding{51} & 44.44 & 44.48\\
&& \cellcolor{green!20}{DeepSeek-V3} & \cellcolor{green!20}Sep. 2024 & \cellcolor{green!20}671 B & \cellcolor{green!20}\ding{51} & \cellcolor{green!20}65.56 & \cellcolor{green!20}\cellcolor{green!20}65.06\\
&& \cellcolor{green!20}{DeepSeek-V3-0324}  &  \cellcolor{green!20}Mar. 2025 & \cellcolor{green!20}685 B & \cellcolor{green!20}\ding{51} & \cellcolor{green!20}70.02 & \cellcolor{green!20}69.21\\
&& \cellcolor{green!20}{QwQ-32B} & Mar. \cellcolor{green!20}2025 & \cellcolor{green!20}32 B & \cellcolor{green!20}\ding{51} & \cellcolor{green!20}65.56 & \cellcolor{green!20}62.71\\
&& \cellcolor{green!20}{Qwen3-32B} & \cellcolor{green!20}May 2025 & \cellcolor{green!20}32 B & \cellcolor{green!20}\ding{51} & \cellcolor{green!20}61.11 & \cellcolor{green!20}55.79\\
&& \cellcolor{green!20}{Qwen3-235B-A22B} & \cellcolor{green!20}May 2025 & \cellcolor{green!20}235 B & \cellcolor{green!20}\ding{51} & \cellcolor{green!20}54.44 & \cellcolor{green!20}50.23\\
&& \cellcolor{green!20}{GLM-Z1-Air} & \cellcolor{green!20}Apr. 2025 & \cellcolor{green!20}32 B & \cellcolor{green!20}\ding{51} & \cellcolor{green!20}68.89 & \cellcolor{green!20}68.16\\

\cmidrule{1-8}
\multirow{28}{*}{\makecell{\textbf{Measured}\\\textbf{HRRP}\\\textbf{Dataset}}}&\multirow{3}{*}{\makecell{\textbf{Traditional}\\\textbf{ML}}} 
& SVM-HRRP~\cite{svm}  & Sep. 2005 & - & \ding{55} & 53.33 & 45.93 \\ 
&& SVM-SC~\cite{svm_sc}  & Dec. 2007& - & \ding{55} & 52.22 & 42.22\\
&& RF-SC~\cite{rf}  & Aug. 2018 & - & \ding{55} & \underline{55.56} & 46.85\\
\cmidrule{2-8}

&\multirow{15}{*}{\makecell{\textbf{API-based}\\\textbf{LLMs}\\\textbf{(Proposed)}}} 
& {GPT-4o-mini}  & Jul. 2024 & 8 B & \ding{51} & 33.33 & 33.06
\\
&& {GPT-3.5-turbo}  & Mar. 2023 & 20 B & \ding{51} & 43.33 & 41.54\\
&& {GPT-4}  & Apr. 2023 & - & \ding{51} & 40.00 & 37.06 \\
&& \cellcolor{green!20}{GPT-4-turbo}  & \cellcolor{green!20}Jan. 2024 & \cellcolor{green!20}- & \cellcolor{green!20}\ding{51} & \cellcolor{green!20}45.56 & \cellcolor{green!20}44.29 \\
&& \cellcolor{green!20}{GPT-o4-mini}  & \cellcolor{green!20}Apr. 2025 & \cellcolor{green!20}- & \cellcolor{green!20}\ding{51} & \cellcolor{green!20}44.44 & \cellcolor{green!20}43.96 \\
&& \cellcolor{green!20}{GPT-4.1}  & \cellcolor{green!20}Apr. 2025 & \cellcolor{green!20}- & \cellcolor{green!20}\ding{51} & \cellcolor{green!20}46.67 & \cellcolor{green!20}44.74\\
&& \cellcolor{green!20}{Claude-3-sonnet}  & \cellcolor{green!20}Feb. 2024 & \cellcolor{green!20}- & \cellcolor{green!20}\ding{51} & \cellcolor{green!20}43.33 & \cellcolor{green!20}42.71\\
&& \cellcolor{green!20}{Claude-3-opus} & \cellcolor{green!20}Feb. 2024 & \cellcolor{green!20}- & \cellcolor{green!20}\ding{51} & \cellcolor{green!20}47.78 & \cellcolor{green!20}45.63 \\
&& \cellcolor{green!20}{Claude-3-5-sonnet} & \cellcolor{green!20}Oct. 2024 & \cellcolor{green!20}- & \cellcolor{green!20}\ding{51} & \cellcolor{green!20}48.89 & \cellcolor{green!20}47.90\\
&& \cellcolor{green!20}{Claude-3-7-sonnet} & \cellcolor{green!20}Feb. 2025 & \cellcolor{green!20}- & \cellcolor{green!20}\ding{51} & \cellcolor{green!20}44.44 & \cellcolor{green!20}43.41 \\
&& \cellcolor{green!20}{Claude-4-sonnet} & \cellcolor{green!20}May 2025 & \cellcolor{green!20}- & \cellcolor{green!20}\ding{51} & \cellcolor{green!20}48.89 & \cellcolor{green!20}46.89\\
&& \cellcolor{green!60}{\textbf{Claude-4-opus}} & \cellcolor{green!60}\textbf{May 2025} & \cellcolor{green!60}\textbf{-} & \cellcolor{green!60}\ding{51} & \cellcolor{green!60}\textbf{57.78} & \cellcolor{green!60}\textbf{56.16}\\
&& \cellcolor{green!20}{Gemini-1.5-Pro} & \cellcolor{green!20}Sep. 2024 & \cellcolor{green!20}175 B & \cellcolor{green!20}\ding{51} & \cellcolor{green!20}50.00 & \cellcolor{green!20}48.29\\
&& {GLM-4-Air} & Jun. 2024 & - & \ding{51} & 35.56 & 24.76\\
&& {GLM-4-Plus} & Sep. 2024 & - & \ding{51} & 38.89 & 37.86\\
\cmidrule{2-8}
&\multirow{10}{*}{\makecell{\textbf{Open-source}\\\textbf{LLMs}\\\textbf{(Proposed)}}}  & \cellcolor{green!20}{DeepSeek-R1-Distill-Qwen-7B}  & \cellcolor{green!20}Jan. 2025 & \cellcolor{green!20}7 B & \cellcolor{green!20}\ding{51} & \cellcolor{green!20}45.56 & \cellcolor{green!20}44.39\\
&& {DeepSeek-R1-Distill-Llama-8B}& Jan. 2025 & 8 B & \ding{51} & 38.89 & 37.86 \\
&& \cellcolor{green!20}{DeepSeek-R1-Distill-Qwen-14B}  & \cellcolor{green!20}Jan. 2025 & \cellcolor{green!20}14 B & \cellcolor{green!20}\ding{51} & \cellcolor{green!20}43.33 & \cellcolor{green!20}43.18\\
&& \cellcolor{green!20}{DeepSeek-R1-Distill-Qwen-32B} & \cellcolor{green!20}Jan. 2025 & \cellcolor{green!20}32 B & \cellcolor{green!20}\ding{51} & \cellcolor{green!20}45.56 & \cellcolor{green!20}45.41\\
&& {DeepSeek-V3} & Sep. 2024 & 671 B & \ding{51} & 42.22 & 41.94\\
&& {DeepSeek-V3-0324}  &  Mar. 2025 & 685 B & \ding{51} & 41.11 & 41.84\\
&& \cellcolor{green!60}{\textbf{QwQ-32B}} & \cellcolor{green!60}{\textbf{Mar. 2025}} & \cellcolor{green!60}{\textbf{32 B}} & \cellcolor{green!60}\ding{51} & \cellcolor{green!60}50.12 & \cellcolor{green!60}\underline{48.75}\\
&& {Qwen3-32B} & May 2025 & 32 B & \ding{51} & 37.78 & 34.14\\
&& {Qwen3-235B-A22B} & May 2025 & 235 B & \ding{51} & 33.33 & 18.61\\
&& \cellcolor{green!20}{GLM-Z1-Air} & \cellcolor{green!20}Apr. 2025 & \cellcolor{green!20}32 B & \cellcolor{green!20}\ding{51} & \cellcolor{green!20}46.67 & \cellcolor{green!20}46.22\\

\hline
\end{tabular}
\begin{tablenotes}
        \small
        \item * The best results are in \textbf{bold}, and the second is \underline{underlined}. LLMs that produce results better than at least 1 metric of the tradition methods, and LLMs that produce the best or second results at any metric are highlighted in \protect\includegraphics[scale=0.99]{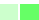} (light green and green), respectively.
      \end{tablenotes}
  \end{threeparttable}
\end{adjustbox}

\end{table}

\begin{figure}[ht]
    \centering
    \includegraphics[width=\linewidth]{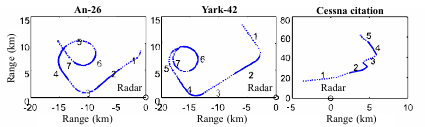}
    \caption{The target trajectories of our 3 types measured dataset.}
    \label{fig:dataset}
\end{figure}

\begin{figure}[ht]
    \centering
    \includegraphics[width=\linewidth]{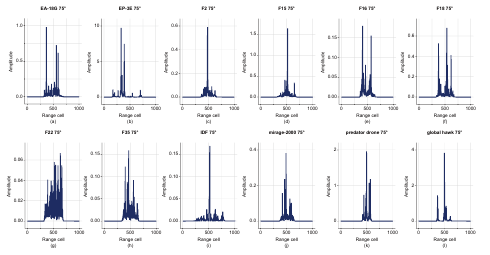}
    \caption{Examples of all 12 types in the simulated dataset at the same aspect.}
    \label{fig:dataset_samples}
\end{figure}

\begin{figure*}[t]
    \centering
    \includegraphics[width=\linewidth]{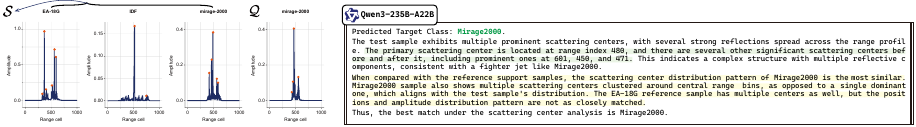} 
    \caption{An example task consists of EA-18G, IDF, and Mirage-2000 (3-way 1-shot), the query is a Mirgae-2000 sample. The right shows the response from Qwen3-235B-A22B. Location-related contents and the comparison-related contents are highlighted in \protect\includegraphics[scale=0.50]{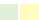} (light green and yellow), respectively.}
\end{figure*}

\begin{figure}[ht]
    \centering
    \includegraphics[width=\linewidth]{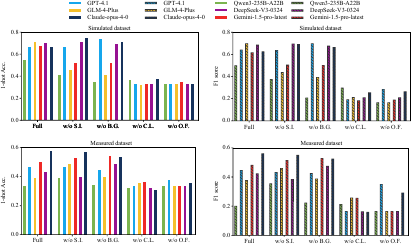} 
    \caption{Impact of prompt components on FSL performance. Each group of bars represent a prompt component ablation (\textit{i.e.}, w/o System Instruction, w/o Background knowledge, w/o Candidate List, and w/o Output Format) compared to the Full Prompt.}
    \label{fig:prompt_ablation_placeholder}
\end{figure}

\begin{figure}[ht]
    \centering
    \includegraphics[width=\linewidth]{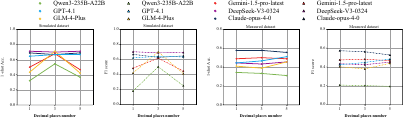} 
    \caption{Impact of SC decimal places of position index and amplitude on FSL performance. Groups show variations 1 \textit{vs.} 3 \textit{vs.} 5.}
    \label{fig:sc_quality_ablation_placeholder}
\end{figure}

\begin{figure}[ht]
    \centering
    \includegraphics[width=\linewidth]{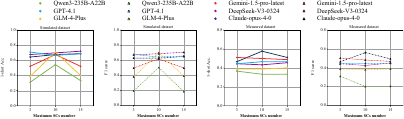} 
    \caption{Impact of SC maximum number on FSL performance. Groups show variations in $N_{sc\_max}$ (\textit{i.e.}, 3 \textit{vs.} 10 \textit{vs.} 15).}
    \label{fig:sc_number_ablation_placeholder}
\end{figure}

\begin{figure}[ht]
    \centering
    \includegraphics[width=\linewidth]{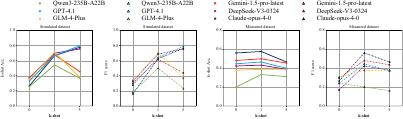} 
    \caption{Impact of number of support shots on FSL performance. Specifically, $K$=0, $K$=1, and $K$=5 are included.}
    \label{fig:kshot_ablation_placeholder}
\end{figure}

\subsection{Experimental Setup}

\subsubsection{Datasets and SC Representation}

Our datasets: \textbf{simulated dataset:} this dataset comprises 12 aircraft classes (\textit{e.g.}, EA-18G, F22, F15), simulated via electromagnetic software. Data were generated in X-band (9.5\textasciitilde10.5 GHz) across varied azimuths ($0^\circ$\textasciitilde$60^\circ$, $0.05^\circ$ step) and pitches ($-15^\circ$\textasciitilde$15^\circ$, $3^\circ$ step) for four polarizations (HH, HV, VH, VV), resulting in $4 \times 11 \times 1201$ profiles per class, each with 984 range cells; \textbf{measured dataset:} a widely used public dataset~\cite{liu2024prior, xu2019targetaware} containing three aircraft types (An-26, Yark-42, Cessna Citation). These C-band radar (5.52 GHz center, 400 MHz bandwidth) measurements yield HRRPs with 306 range cells per sample. For each HRRP signal, SCs are extracted using a peak detection method with consistent parameters: prominence of 0.15, min distance between peaks of 5, and $N_{sc\_max}=10$. Figure \ref{fig:dataset} shows a detailed the tarjectories of our measured dataset. In our experiments, we chose the measured data from the ``1'' part of the trajectories to form measured dataset.

\subsubsection{Platform}
We conducted all our experiments on a Linux server with 3*NVIDIA A4000 GPUs, 2*Intel Xeon Bronze CPUs, and 64GB RAM, yet our training-free HRRPLLM don't need a CUDA environment, APIs are public-available.

\subsubsection{LLMs and Baselines}
We evaluate a thorough set of \textbf{latest} LLMs (till May 2025), with various architectures, sizes, and providers. For Google Gemini series, we failed to achieve API call due to their block of contents like ``F22'' \textit{etc}. Several baselines were compared against: traditional ML methods include Support Vector Machines (SVM) trained on either HRRP amplitudes (SVM-HRRP) or SCs (SVM-SC)~\cite{svm, svm_sc}, and Random Forest (RF-SC)~\cite{rf}. For a fair comparison, we did not include current FSL methods because their need of meta-training datasets and no public-available testing weights.

\subsection{Comparative Results and Analysis}
\label{ssec:llm_results}

The performance of our proposed HRRPLLM framework on different LLMs is compared against traditional ML baselines. The LLMs and traditional ML baselines are similarly evaluated on 3-way 1-shot tasks. Traditional ML baselines are trained on the support set and evaluated on the query, while our LLMs takes the whole task as prompt input and directly output ATR results. Table~\ref{tab:comparative_results} summarizes these results.  Specifically, Claude-4-sonnet achieves {75.56\%} 1-shot Acc. and {75.40\%} F1. on simulated dataset, respectively outperforms baselines {24.45\%} and {31.51\%} at maximum. On the measured dataset where the aspect variation is more significant, our HRRPLLM models still show better performance. For 1-shot Acc. and F1., Claude-4-opus outperforms baselines at largest {5.56\%} and {13.94\%}.

\subsection{Ablation Studies}
\label{ssec:ablation}

\subsubsection{Impact of Prompt Components}
We investigate the necessity of components within the prompt structure. This includes: \textbf{(1) S.I.:} system instruction and role definition; \textbf{(2)  B.G.:} background knowledge for SC characteristics; \textbf{(3) C.L.:} list of candidate target types; and \textbf{(4) O.F.:} output format. The results in Fig.~\ref{fig:prompt_ablation_placeholder} indicate that S.I. and B.G. indeed boost the performance of LLMs, while lack of C.I. or O.F. can be fatal.

\subsubsection{Influence of SCs Presentation}
The quality and quantity of the input scattering center information are critical. On the one hand, We study this by varying $N_{sc\_max}$ during SC extraction. Additionally, we explore the effect of altering the numerical precision of the encoded SC amplitudes. As Fig.~\ref{fig:sc_quality_ablation_placeholder}\textasciitilde\ref{fig:sc_number_ablation_placeholder} demonstrates, neither increasing nor decreasing $N_{sc\_max}$ or decimal places of SCs benefit the performance. 


\subsubsection{Effect of Few-shot Support Samples}
To quantify the benefit of in-context learning from support examples, we explicitly compare the performance when $K$=0, where only the task context and query are provided against scenarios with $K$=1 and $K$=5 support samples. Results in Fig.~\ref{fig:kshot_ablation_placeholder} reveals that increasing $K$ might promote performance. if the model's context window is big (\textit{e.g.}, Claude-opus-4.0, GPT-4.1). However, the influence of aspect variation can be greater, leading to declined results on the measured dataset.

\section{Conclusion}

This letter introduces HRRPLLM, a framework revealing the fact that current general-purpose LLMs can achieve competitive, explainable, and training-free few-shot HRRP ATR. These abilities are crucial for rapid adaptation to few-shot novel targets and trustworthy AI via prediction reasons. Current limitations, \textit{e.g.} aspect sensitivity and API reliance, motivate promising future researches into aspect-specific LLM prompt-tuning and efficient local deployment.




\newpage
\begin{thebibliography}{99}
\bibitem{li2025saratrx}
W. Li, W. Yang, Y. Hou, L. Liu, Y. Liu, and X. Li, ``SARATR-X: Toward Building a Foundation Model for SAR Target Recognition,'' \textit{IEEE Trans. Image Process.}, vol. 34, pp. 869--884, 2025.

\bibitem{yang2024radar}
L. Yang, W. Feng, Y. Wu, L. Huang, and Y. Quan, ``Radar-Infrared Sensor Fusion Based on Hierarchical Features Mining,'' \textit{IEEE Signal Process. Lett.}, vol. 31, pp. 66--70, 2024.

\bibitem{radford2021learning}
A. Radford, J. W. Kim, C. Hallacy, A. Ramesh, G. Goh, S. Agarwal, G. Sastry, A. Askell, P. Mishkin, J. Clark, G. Krueger, and I. Sutskever, ``Learning Transferable Visual Models From Natural Language Supervision,'' in \textit{Proc. 38th Int. Conf. Mach. Learn. (ICML)}, 2021, pp. 8748--8763.

\bibitem{chen2024hrrpgraphnet}
L. Chen, X. Sun, Z. Pan, Q. Liu, Z. Wang, X. Sy, Z. Liu, and P. Hu, ``HRRPGraphNet: Make HRRPs to be graphs for efficient target recognition,'' \textit{Electron. Lett.}, vol. 60, no. 22, Art. no. e70088, 2024.

\bibitem{liu2024prior}
Q. Liu, X. Zhang, and Y. Liu, ``A Prior-Knowledge-Guided Neural Network Based on Supervised Contrastive Learning for Radar HRRP Recognition,'' \textit{IEEE Trans. Aerosp. Electron. Syst.}, vol. 60, no. 3, pp. 2854--2873, Jun. 2024.

\bibitem{liu2021gafmn}
Q. Liu, X. Zhang, and Y. Liu, ``GAF-MN: A New HRRP Target Recognition Method Based on Gramian Angular Field and Matching Networks in Few-Shot Condition," in \textit{Proc. CIE Int. Conf. Radar (Radar 2021)}, Dec. 2021, pp. 1288--1292.

\bibitem{liu2023contrastive}
M. Liu, Z. Zhang, and X. Gao, ``Contrastive learning-based prototype calibration method for few-shot HRRP recognition,'' in \textit{Proc. IET Int. Radar Conf. (IRC 2023)}, Dec. 2023, pp. 920--923.

\bibitem{liu2021fewshot}
Q. Liu, X. Zhang, Y. Liu, K. Huo, W. Jiang, and X. Li, ``Few-Shot HRRP Target Recognition Based on Gramian Angular Field and Model-Agnostic Meta-Learning,'' in \textit{Proc. IEEE 6th Int. Conf. Signal Image Process. (ICSIP)}, Oct. 2021, pp. 6--10.

\bibitem{liu2024remoteclip}
F. Liu, D. Chen, Z. Guan, X. Zhou, J. Zhu, Q. Ye, L. Fu, and J. Zhou, ``RemoteCLIP: A Vision Language Foundation Model for Remote Sensing,'' \textit{IEEE Trans. Geosci. Remote Sens.}, vol. 62, pp. 1--16, 2024.

\bibitem{wang2023type}
Y. Wang, Y. Ma, Z. Zhang, X. Zhang, and L. Zhang, ``Type-Aspect Disentanglement Network for HRRP Target Recognition With Missing Aspects,'' \textit{IEEE Geosci. Remote Sens. Lett.}, vol. 20, pp. 1--5, 2023.

\bibitem{li2024priorinfo}
J. Li, W. Guo, F. Wei, T. Zhang, and W. Yu, ``Prior Information-Assisted Few-Shot HRRP Recognition Based on Task-Wise Shrinkage Quadratic Discriminant Analysis,'' \textit{IEEE Trans. Aerosp. Electron. Syst.}, vol. 60, no. 6, pp. 9354--9368, Dec. 2024.

\bibitem{song2022multiview}
Y. Song, Q. Zhou, W. Yang, Y. Wang, C. Hu, and X. Hu, ``Multi-View HRRP Generation With Aspect-Directed Attention GAN,'' \textit{IEEE J. Sel. Topics Appl. Earth Observ. Remote Sens.}, vol. 15, pp. 7643--7656, 2022.

\bibitem{guo2025hrrp}
Z. Guo, Z. Liu, R. Xie, and L. Ran, ``HRRP Few-Shot Target Recognition for Full Polarimetric Radars via SCs Optimal Matching,'' \textit{IEEE Trans. Aerosp. Electron. Syst.}, vol. 61, no. 2, pp. 4526--4541, Apr. 2025.

\bibitem{liu2025attribute}
Q. Liu, X. Zhang, and Y. Liu, ``Attribute-Informed and Similarity-Enhanced Zero-Shot Radar Target Recognition,'' \textit{IEEE Trans. Aerosp. Electron. Syst.}, preprint, pp. 1--19, 2025.

\bibitem{liu2025radarhybrid}
X. Liu, D. Zhou, and Q. Huang, ``Radar HRRP Target Recognition Based on Hybrid Quantum Neural Networks,'' \textit{IEEE Trans. Aerosp. Electron. Syst.}, pp. 1--16, 2025.

\bibitem{liu2024scnet}
Q. Liu, X. Zhang, and Y. Liu, ``SCNet: Scattering center neural network for radar target recognition with incomplete target-aspects,'' \textit{Signal Process.}, vol. 219, no. C, Jun. 2024.

\bibitem{xia2023radaropen}
Z. Xia, P. Wang, G. Dong, and H. Liu, ``Radar HRRP Open Set Recognition Based on Extreme Value Distribution,'' \textit{IEEE Trans. Geosci. Remote Sens.}, vol. 61, pp. 1--16, 2023.

\bibitem{liu2021multipolarization}
Q. Liu, X. Zhang, Y. Liu, K. Huo, W. Jiang, and X. Li, ``Multi-Polarization Fusion Few-Shot HRRP Target Recognition Based on Meta-Learning Framework," \textit{IEEE Sensors J.}, vol. 21, no. 16, pp. 18085--18100, Aug. 2021.

\bibitem{li2025mtbc}
S. Li, W. Li, P. Huang, M. Zheng, B. Tian, and S. Xu, ``MTBC: Masked Vision Transformer and Brownian Distance Covariance Classifier for Cross-Domain Few-Shot HRRP Recognition,'' \textit{IEEE Sensors J.}, vol. 25, no. 9, pp. 16440--16454, May 2025.

\bibitem{sagayaraj2021comparative}
M. J. Sagayaraj, V. Jithesh, and D. Roshani, ``Comparative Study Between Deep Learning Techniques and Random Forest Approach for HRRP Based Radar Target Classification,'' in \textit{Proc. Int. Conf. Artif. Intell. Smart Syst. (ICAIS 2021)}, Mar. 2021, pp. 385--388.

\bibitem{wang2024tabletime}
J. Wang, M. Cheng, Q. Mao, Q. Liu, F. Xu, X. Li, and E. Chen, ``TableTime: Reformulating Time Series Classification as Zero-Shot Table Understanding via Large Language Models,'' 2024, \textit{arXiv:2411.15737}.

\bibitem{dong2024survey}
Q. Dong, L. Li, D. Dai, C. Zheng, J. Ma, R. Li, H. Xia, J. Xu, Z. Wu, B. Chang, X. Sun, L. Li, and Z. Sui, ``A Survey on In-context Learning,'' in \textit{Proc. Conf. Empirical Methods Natural Lang. Process. (EMNLP 2024)}, Nov. 2024, pp. 1107--1128.

\bibitem{liu2022whatmakes}
J. Liu, D. Shen, Y. Zhang, B. Dolan, L. Carin, and W. Chen, ``What Makes Good In-Context Examples for GPT-3?,'' in \textit{Proc. Deep Learn. Inside Out (DeeLIO 2022)}, May 2022, pp. 100--114.

\bibitem{liu2024infmae}
F. Liu, C. Gao, Y. Zhang, J. Guo, J. Wang, and D. Meng, ``InfMAE: A Foundation Model in the Infrared Modality,'' in \textit{Proc. Eur. Conf. Comput. Vis. (ECCV 2024)}, Sep. 2024, pp. 420--437.

\bibitem{tang2025time}
H. Tang, C. Zhang, M. Jing, Q. Yu, Z. Wang, X. Jin, Y. Zhang, and M. Du, ``Time Series Forecasting with LLMs: Understanding and Enhancing Model Capabilities,'' \textit{SIGKDD Explor. Newsl.}, vol. 26, no. 2, pp. 109--118, Jan. 2025.

\bibitem{gruver2023large}
N. Gruver, M. Finzi, S. Qiu, and A. G. Wilson, ``Large language models are zero-shot time series forecasters,'' in \textit{Proc. 37th Int. Conf. Neural Inf. Process. Syst. (NeurIPS)}, Dec. 2023, pp. 19622--19635.

\bibitem{jiang2024empowering}
Y. Jiang, Z. Pan, X. Zhang, S. Garg, A. Schneider, Y. Nevmyvaka, and D. Song, ``Empowering time series analysis with large language models: a survey,'' in \textit{Proc. 33rd Int. Joint Conf. Artif. Intell. (IJCAI)}, Aug. 2024, pp. 8095--8103.

\bibitem{zhang2025enhancing}
W. Zhang, Z. Cheng, F. Lei, B. Liu, D. Gu, and Z. Wang, ``Enhancing Logical Reasoning in Large Language Models via Multi-Stage Ensemble Architecture with Adaptive Attention and Decision Voting,'' in \textit{Proc. 5th Int. Conf. Big Data Econ. Inf. Manag. (BDEIM)}, May 2025, pp. 1201--1205.

\bibitem{xu2019targetaware}
B. Xu, B. Chen, J. Wan, H. Liu, and L. Jin, ``Target-Aware Recurrent Attentional Network for Radar HRRP Target Recognition,'' \textit{Signal Process.}, vol. 155, pp. 268--280, Feb. 2019.

\bibitem{li2023fewshotigarss}
J. Li, D. Li, Y. Jiang, and W. Yu, ``Few-Shot Radar HRRP Recognition Based on Improved Prototypical Network," in \textit{Proc. IEEE Int. Geosci. Remote Sens. Symp. (IGARSS 2023)}, 2023, pp. 5277-5280.

\bibitem{kumar_meta-learning_2024}
S. Kumar, A. Sharma, V. Shokeen, A. T. Azar, S. U. Amin, and Z. I. Khan, ``Meta-learning for real-world class incremental learning: a transformer-based approach," \textit{Sci. Rep.}, vol. 14, no. 1, p. 23092, Oct. 2024.

\bibitem{wolpert1997nofree}
D.H. Wolpert and W.G. Macready, ``No free lunch theorems for optimization,'' \textit{IEEE Trans. Evol. Comput.}, vol. 1, no. 1, pp. 67-82, 1997.

\bibitem{li2024sardet100k}
Y. Li, X. Li, W. Li, Q. Hou, L. Liu, M. Chen, and J. Yang, ``SARDet-100K: Towards Open-Source Benchmark and ToolKit for Large-Scale SAR Object Detection,'' in \textit{Proc. 38th Int. Conf. Neural Inf. Process. Syst. (NeurIPS), 2024}

\bibitem{finn2017modelagnostic}
C. Finn, P. Abbeel, and S. Levine, ``Model-agnostic meta-learning for fast adaptation of deep networks,'' in \textit{Proc. 34th Int. Conf. Mach. Learn. (ICML)}, 2017, pp. 1126--1135.

\bibitem{vinyals2016matching}
O. Vinyals, C. Blundell, T. Lillicrap, K. Kavukcuoglu, and D. Wierstra, ``Matching networks for one shot learning,'' in \textit{Proc. 30th Int. Conf. Neural Inf. Process. Syst. (NIPS), 2016}, pp. 3637--3645.

\bibitem{snell2017prototypical}
J. Snell, K. Swersky, and R. Zemel, ``Prototypical networks for few-shot learning,'' in \textit{Proc. 31st Int. Conf. Neural Inf. Process. Syst. (NIPS)}, 2017, pp. 4080--4090.

\bibitem{svm}
X. Wang and C. Wu, ``Using improved SVM decision tree to classify HRRP,'' in \textit{Proc. IEEE Int. Conf. Mach. Learn. Cybern. (ICMLC 2005)}, 2005, pp. 4432-4436.

\bibitem{rf}
Y. Wang, W. Chen, J. Song, Y. Li and X. Yang, ``Open Set Radar HRRP Recognition Based on Random Forest and Extreme Value Theory,'' in \textit{Proc. IEEE Int. Conf. Radar (RADAR 2018)}, 2018, pp. 1-4.,

\bibitem{svm_sc}
H. Liang and C. Tong, ``SVM Target Identification Method Based on HRRP Sample Partition,'' in \textit{Proc. IEEE Asia-Pac. Microw. Conf. (APMC 2007)}, 2007, pp. 1-4.

\end{thebibliography}
\end{document}